\documentclass[a4paper]{jpconf}
\usepackage{graphicx}
\begin{document}
\title{Correlations of the stress-energy tensor in AdS spaces via the generalized zeta-function method}

\author{H. T. Cho$^{1}$ and B. L. Hu$^{2}$}

\address{$^{1}$ Department of Physics, Tamkang University, Tamsui, New Taipei City, TAIWAN\\
$^{2}$ Maryland Center for Fundamental Physics, Department of Physics, University of Maryland, College Park, Maryland 20742-4111, USA}

\ead{$^{1}$htcho@mail.tku.edu.tw ; $^{2}$blhu@umd.edu}

\begin{abstract}
We calculate the vacuum expectation values of the stress-energy bitensor of a minimally coupled massless scalar field in anti-de Sitter (AdS) spaces. These correlators, also known as the noise kernel, act as sources in the Einstein-Langevin equations of stochastic gravity \cite{HuVer08,HuVer03} which govern the induced metric fluctuations beyond the mean-field dynamics described by the semiclassical Einstein equations of semiclassical gravity. Because the AdS spaces are  maximally symmetric the eigenmodes have analytic expressions which facilitate the computation of the  zeta-function \cite{DowCri76,Haw77}. Upon taking the second functional variation of the generalized  zeta function introduced in \cite{PH97} we obtain  the correlators of the stress tensor.  Both the short and the long geodesic distance limits of the correlators are presented.
\end{abstract}

\section{Introduction}

After the seminal work of Hawking \cite{Haw75} it is realized that with quantum effects black holes can evaporate. Black holes behave like thermal objects and can be in equilibrium with a thermal gas at the Hawking temperature. However, this equilibrium state is unstable to the absorption of the thermal gas by the black hole in an asymptotically flat spacetime \cite{Haw76}. In order to obtain an equilibrium state one has to put the black hole in an ad hoc cavity with some appropriate boundary condition \cite{Yor85}. The situation with black holes in an anti-de Sitter (AdS) spacetime is rather different as explored by Hawking and Page \cite{HP83}. Above the critical temperature there are two masses of the black hole which can be in equilibrium with the thermal gas. The larger mass black hole is actually stable. This phase transition is now known as the Hawking-Page transition.

The recent interest in this Hawking-Page transition is mainly due to the AdS/CFT correspondence \cite{Wit98}. This transition is conjectured to be related to the confinement phenomenon in QCD under this correspondence. Moreover, the quasinormal modes of the AdS black hole perturbations are supposed to give information on the thermal states of the conformal theory at the boundary \cite{HorHub}.

To study the Hawking-Page transition in a dynamical manner, the corresponding fluctuations of the quantum fields on the background spacetime must be taken into account. A systematic way to do so is by the theory of stochastic gravity \cite{HuVer08,HuVer03}. To the lowest order the fluctuations are represented by a noise kernel which is the vacuum expectation value of the correlation of the stress energy tensor. This is one of the reasons why we are interested in such a consideration here. For the AdS black hole case, one should deal with the stress tensor of a thermal gas in the Hartle-Hawking vacuum state. However, before taking such an endeavor we would like to get a grip on the problem by calculating the case with the scalar field in AdS spacetime.

The stress energy tensor correlators have been obtained before, notably, by Osborn and Shore
\cite{OsbSho00} for maximally symmetric spacetimes which include the de Sitter and anti-de Sitter spaces via conformal field theory techniques. They have considered both the scalar and the fermion cases. However, they are only interested in the conformally coupled fields and have made extensive use of the traceless condition of the correlator. In \cite{PRVdS} the authors have also calculated this quantity for quantum scalar fields in de Sitter spacetime. Their results can be analytically extended to AdS spaces and vice versa.

As this quantity is of basic importance it is useful to derive these expressions with a different method. The zeta function method we adopt here was introduced by Dowker and Critchley \cite{DowCri76} and Hawking \cite{Haw77}. It is one of the most elegant methods used for the regularization of the stress energy tensor for quantum fields in Riemannian (not pseudo-Riemannian)  spaces. One can use the zeta function to construct the regularized effective action and upon taking its first functional variation obtain the regularized stress energy tensor. (For a systematic exposition, see \cite{Elizalde}). This method is generalized by  Phillips and Hu \cite{PH97} to calculate the noise kernels of quantum scalar fields in $S^1 \times R^1$ spaces (useful for finite temperature theory and Casimir energy density considerations ) and for an Einstein universe, by way of the second order variation of the regularized effective action (see also \cite{CogEli02}). This method is used in our present calculation of the noise kernel for AdS spacetimes in N-dimensions.

In the next section, we shall describe the generalized zeta-function method of Phillips and Hu in some details as applied to a scalar field. Then in Section III, the correlators of the stress-energy tensors in $AdS^{N}$ are calculated. The small and large geodesic distance expansions of the correlators are given in Section IV. The results are then compared with those in \cite{PRVdS}. Finally conclusions and discussions are presented in Section V.

\section{Scalar field in spacetimes admitting an Euclidean section}

To make the discussions general we first consider a massive $m$ scalar field $\phi$ coupled to an $N$-dimensional Euclideanized space (with contravariant metric $g^{\mu\nu}(x)$, determinant $g$ and scalar curvature $R$) with coupling constant $\xi$ described by the action
\begin{eqnarray}
S[\phi]=\frac{1}{2}\int d^{N}x\sqrt{g(x)}\phi(x)H\phi(x),
\end{eqnarray}
where $H$ is the quadratic operator
\begin{eqnarray}
H=-g^{\mu\nu}\nabla_{\mu}\nabla_{\nu}+m^{2}+\xi R,
\end{eqnarray}
and $R$ is the scalar curvature. The effective action defined by $W={\rm ln}{\cal Z}$ is related to
the generating functional ${\cal Z}$ by 
\begin{eqnarray}
{\cal Z}=\int{\cal D}\phi\ \! e^{-S[\phi]}.
\end{eqnarray}
The expectation value of the stress-energy tensor can be obtained by taking the functional derivative of the effective action
\begin{eqnarray}
\langle T_{\mu\nu}\rangle=-\frac{2}{\sqrt{g(x)}}\frac{\delta W}{\delta g^{\mu\nu}(x)}.
\end{eqnarray}
This formal expression is divergent at coincident limit and some procedure of regularization must be implemented. Here we adopt the procedure of $\zeta$-function regularization as shown by Dowker and Critchley \cite{DowCri76} and Hawking \cite{Haw77}. As shown by Phillips and Hu \cite{PH97} the fluctuation of the stress-energy tensor is obtained by taking another derivatives of the regularized effective action $W$,
\begin{eqnarray}
\Delta T^{2}_{\mu\nu\alpha'\beta'}(x,x')&\equiv&\langle T_{\mu\nu}(x)T_{\alpha'\beta'}(x')\rangle-\langle T_{\mu\nu}(x)\rangle\langle T_{\alpha'\beta'}(x')\rangle\nonumber\\
&=&\frac{4}{\sqrt{g(x)g(x')}}\frac{\delta^{2}W}{\delta g^{\mu\nu}(x)\delta g^{\alpha'\beta'}(x')}
\end{eqnarray}
Note that although the regularized expectation value of the stress-energy tensor is dependent on one spacetime point, the fluctuation of the stress energy is a  bitensor defined at two separate spacetime points through the two functional derivatives taken with respect to these two separate spacetime points.

Conventionally one define the $\zeta$-function of an operator $H$,
\begin{eqnarray}
\zeta_{H}(s)=\sum_{n}\left(\frac{\mu}{\lambda_{n}}\right)^{s}={\rm Tr}\left(\frac{\mu}{H}\right)^{s}
\end{eqnarray}
where $\lambda_{n}$ is the eigenvalues of $H$ and $\mu$ represents the renormalization scale.
The $\zeta$-function regularized effective action of the operator $H$ is
\begin{eqnarray}
W_{R}=\left.\frac{1}{2}\frac{d\zeta}{ds}\right|_{s\rightarrow 0}.
\end{eqnarray}
Using the proper-time method \cite{DowCri76}  one can write the $\zeta$-function as
\begin{eqnarray}
\zeta_{H}(s)&=&\frac{\mu^{s}}{\Gamma(s)}\int_{0}^{\infty}dt\ \! t^{s-1}{\rm Tr}e^{-tH}\\
W_{R}&=&\frac{1}{2}\frac{d}{ds}\left[\frac{\mu^{s}}{\Gamma(s)}\int_{0}^{\infty}dt\ \! t^{s-1}{\rm Tr}e^{-tH}\right]_{s\rightarrow 0}
\end{eqnarray}
Taking the first variation of the $\zeta$-function,
\begin{eqnarray}
\delta\zeta_{H}&=&-\frac{\mu^{s}}{\Gamma(s)}\int_{0}^{\infty}dt\ \! t^{s}{\rm Tr}\left(\delta He^{-tH}\right)\nonumber\\
&=&-\frac{\mu^{s}}{\Gamma(s)}\int_{0}^{\infty}dt\ \! t^{s}\sum_{n}e^{-t\lambda_{n}}\langle n\left|\delta H\right|n\rangle
\end{eqnarray}
we obtain the regularized expectation value of the stress-energy tensor which is given by
\begin{eqnarray}
\langle T_{\mu\nu}(x)\rangle&=&\frac{1}{2}\frac{d}{ds}\left[-\frac{\mu^{s}}{\Gamma(s)}\int_{0}^{\infty}dt\ \! t^{s}\sum_{n}e^{-t\lambda_{n}}\left(-\frac{2}{\sqrt{g(x)}}\left\langle n\right|\frac{\delta H}{\delta g^{\mu\nu}(x)}\left|n\right\rangle\right)\right]_{s\rightarrow 0}\nonumber\\
&=&-\frac{1}{2}\frac{d}{ds}\left\{\frac{\mu^{s}}{\Gamma(s)}\int_{0}^{\infty}dt\ \! t^{s}\sum_{n}e^{-t\lambda_{n}}\, T_{\mu\nu}\left[\phi_{n}(x),\phi_{n}^{*}(x)\right]\right\}_{s\rightarrow 0}\label{renstress}
\end{eqnarray}
where
\begin{eqnarray}
T_{\mu\nu}\left[\phi_{n}(x),\phi_{n'}^{*}(x)\right]&\equiv&-\frac{2}{\sqrt{g(x)}}\left\langle n'\right|\frac{\delta H}{\delta g^{\mu\nu}(x)}\left|n\right\rangle\nonumber\\
&=&-\frac{2}{\sqrt{g(x)}}\int d^{N}x'\sqrt{g(x')}\phi_{n'}^{*}(x')\left[\frac{\delta H}{\delta g^{\mu\nu}(x)}\phi_{n}(x')\right]
\end{eqnarray}
and $\phi_{n}(x)$ is a eigenfunction of the operator $H$, namely,
\begin{equation}
H \phi_n = \lambda_n \phi_n
\end{equation}
where $ \lambda_n $ are the eigenvalues corresponding to the eigenfunctions $\phi_n$.
We then have
\begin{eqnarray}
T_{\mu\nu}\left[\phi_{n}(x),\phi_{n'}^{*}(x)\right]&=&
-\left(\partial_{\mu}\phi_{n'}^{*}\partial_{\nu}\phi_{n}+\partial_{\nu}\phi_{n'}^{*}\partial_{\mu}\phi_{n}\right)
+g_{\mu\nu}\left(g^{\alpha\beta}\partial_{\alpha}\phi_{n'}^{*}\partial_{\beta}\phi_{n}
+\phi_{n'}^{*}g^{\alpha\beta}\nabla_{\alpha}\nabla_{\beta}\phi_{n}\right)\nonumber\\
&&\ \ \ -2\xi\left[g_{\mu\nu}g^{\alpha\beta}\nabla_{\alpha}\nabla_{\beta}(\phi_{n'}^{*}\phi_{n})
-\nabla_{\mu}\nabla_{\nu}(\phi_{n'}^{*}\phi_{n})+R_{\mu\nu}\phi_{n'}^{*}\phi_{n}\right]
\end{eqnarray}

In the approach of Phillips and Hu \cite{PH97}, through the use of the Schwinger method \cite{Sch51}, the second variation of the $\zeta$-function can be written as
\begin{eqnarray}
\delta_{2}\delta_{1}\zeta_{H}&=&\frac{\mu^{s}}{2\Gamma(s)}\int_{0}^{\infty}du\int_{0}^{\infty}dv(u+v)^{s}(uv)^{\nu}\nonumber\\
&&\ \ \ \ \ \left\{{\rm Tr}\left[(\delta_{1}H) e^{-uH}(\delta_{2}H) e^{-vH}\right]+{\rm Tr}\left[(\delta_{2}H) e^{-uH}(\delta_{1}H) e^{-vH}\right]\right\}
\end{eqnarray}
and
\begin{eqnarray}
\Delta T_{\mu\nu\alpha'\beta'}^{2}(x,x')&=&\frac{1}{2}\frac{d}{ds}\left\{\frac{\mu^{s}}{\Gamma(s)}
\int_{0}^{\infty}du\int_{0}^{\infty}dv(u+v)^{s}(uv)^{\nu}\sum_{n,n'}e^{-u\lambda_{n}-v\lambda_{n'}}\right.\nonumber\\
&&\ \ \ \ \ \ \left. T_{\mu\nu}\left[\phi_{n}(x),\phi_{n'}^{*}(x)\right]T_{\alpha'\beta'}\left[\phi_{n'}(x'),\phi_{n}^{*}(x')\right]\right.\!{\Bigg \}}_{s,\nu\rightarrow 0}.\label{T2exp}
\end{eqnarray}
Note that in this Phillips-Hu prescription \cite{PH97}, an additional regularization factor $(uv)^{\nu}$ has been introduced. This is because the authors were interested in the fluctuations of the stress-energy tensor, that is, in the coincident limit of $\Delta T_{\mu\nu\alpha'\beta'}^{2}(x,x')$ where under this limit further divergences occur which call for an additional regularization factor. (See also \cite{CogEli02}). However,  our present purpose is focused on getting the correlators with two points separated, i.e., in the non-coincident case. Hence, apart from the fact that the expression in Eq.~(\ref{T2exp}) is more symmetric with this factor, the keeping of this factor above is actually a matter of convenience. Here we can first take the $s\rightarrow 0$ limit without spoiling the regularization and the expression in Eq.~(\ref{T2exp}) will become
\begin{eqnarray}
\Delta T_{\mu\nu\alpha'\beta'}^{2}(x,x')&=&\frac{1}{2}
\int_{0}^{\infty}du\int_{0}^{\infty}dv (uv)^{\nu}\sum_{n,n'}e^{-u\lambda_{n}-v\lambda_{n'}}\nonumber\\
&&\ \ \ \ \ \ T_{\mu\nu}\left[\phi_{n}(x),\phi_{n'}^{*}(x)\right]T_{\alpha'\beta'}
\left[\phi_{n'}(x'),\phi_{n}^{*}(x')\right]\!{\bigg |}_{\nu\rightarrow 0}.\label{T2finalexp}
\end{eqnarray}
We shall see from the following calculations that with this expression the integrations over $u$ and $v$ effectively separate. The calculations are therefore simplified considerably.

\section{Correlations of the stress-energy tensor}

The Euclideanized N-dim AdS space or the hyperbolic space $H^{N}$ can be described by the metric
\begin{eqnarray}
ds^{2}=d\sigma^{2}+a^{2}\sinh^{2}\left(\frac{\sigma}{a}\right)d\Omega_{N-1}^{2}
\end{eqnarray}
where $\sigma$ is the geodesic distance,
$a$ the radius of the $AdS^N$ space, and $d\Omega_{N-1}^{2}$ the metric for the $(N-1)$-sphere. The eigenfunctions $\phi_{\kappa lm}$ obey the equations,
\begin{eqnarray}
\left(-g^{\mu\nu}\nabla_{\mu}\nabla_{\nu}-\frac{\rho_{N}}{a^{2}}\right)\phi_{\kappa lm}=\left(\frac{\kappa^{2}}{a^{2}}\right)\phi_{\kappa lm}
\end{eqnarray}
where $\rho_{N}=(N-1)/2$, and are given by
\begin{eqnarray}
\phi_{\kappa lm}&=&c_{l}(\kappa)\left(\sinh\frac{\sigma}{a}\right)^{1-\frac{N}{2}}
P_{-\frac{1}{2}+i\kappa}^{1-l-\frac{N}{2}}\left(\cosh\frac{\sigma}{a}\right)Y_{lm}(\Omega)\nonumber\\
&\equiv&f_{\kappa l}(\sigma)Y_{lm}(\Omega)
\end{eqnarray}
where $P_{\nu}^{\mu}(x)$ is the associated Legendre function. $Y_{lm}(\Omega)$ are the hyperspherical harmonics obeying the addition theorem,
\begin{eqnarray}
\sum_{m}Y^{*}_{lm}(\Omega)Y_{lm}(\Omega')=\frac{(2l+N-2)\Gamma\left(\frac{N-2}{2}\right)}{4\pi^{\frac{N}{2}}}
C_{l}^{\frac{N-2}{2}}\left(\Omega\cdot\Omega'\right)\label{addtheorem}
\end{eqnarray}
where $C_{l}^{n}$ is the Gegenbauer polynomial.

The normalization constant is given by
\begin{eqnarray}
c_{l}(\kappa)=\frac{|\Gamma(i\kappa+\rho_{N})|}{|\Gamma(i\kappa)|}
\end{eqnarray}
Later we shall need the normalization constant for $l=0$. For odd $N$,
\begin{eqnarray}
|c_{0}(\kappa)|^{2}=\frac{1}{a^{N}}\prod_{j=0}^{\rho_{N}}\left(\kappa^{2}+j^{2}\right)
=\frac{1}{a^{N}}\sum_{n=1}^{\rho_{N}}c_{2n}\kappa^{2n}\label{c0odd}
\end{eqnarray}
and for even $N$,
\begin{eqnarray}
|c_{0}(\kappa)|^{2}=\frac{1}{a^{N}}(\kappa\tanh\pi\kappa)\prod_{j=\frac{1}{2}}^{\rho_{N}}\left(\kappa^{2}+j^{2}\right)
=\frac{1}{a^{N}}\tanh\pi\kappa\sum_{n=0}^{\rho_{N}-\frac{1}{2}}c_{2n+1}\kappa^{2n+1}\label{c0even}
\end{eqnarray}
In both cases we have turned it into a finite sum.

In a maximally symmetric space like the hyperbolic space $H^{N}$, any bitensor can be expressed in terms of a set of basic bitensors \cite{AllJac86}. The first basic bitensor is the bi-scalar function $\tau(x,x')$
which is the geodesic distance between $x$ and $x'$. Using the covariant derivative one could define
\begin{equation}
n_{\mu}=\nabla_{\mu}\tau(x,x')\ \ \ ;\ \ \ n_{\alpha'}=\nabla_{\alpha'}\tau(x,x')
\end{equation}
where $n_{\mu}(x,x')$ is a vector at $x$ and a scalar at $x'$, while $n_{\alpha'}(x,x')$ is a scalar at $x$ and a vector at $x'$. Next, we have the parallel propagator ${g_{\mu}}^{\alpha'}(x,x')$ which parallel transports any vector $v^{\mu}$ from $x$ to $x'$. The transported vector is $v^{\mu}{g_{\mu}}^{\alpha'}(x,x')$. It is easy to see that $n_{\mu}(x,x')=-{g_{\mu}}^{\alpha'}n_{\alpha'}(x,x')$. $\tau(x,x')$, $n_{\mu}(x,x')$, $n_{\alpha'}(x,x')$ and ${g_{\mu}}^{\alpha'}(x,x')$ constitute this set of basic bitensors. All other bitensors can be expressed in terms of them with coefficients depending only on the geodesic distance between the two points. For example,
\begin{eqnarray}
\nabla^{\mu}n_{\nu}&=&A(\tau)\left({g^{\mu}}_{\nu}-n^{\mu}n_{\nu}\right)\label{der1}\\
\nabla^{\mu}n_{\alpha'}&=&B(\tau)\left({g^{\mu}}_{\alpha'}+n^{\mu}n_{\alpha'}\right)\label{der2}\\
\nabla^{\mu}g_{\nu\alpha'}&=&-(A(\tau)+B(\tau))\left({g^{\mu}}_{\nu}n_{\alpha'}+{g^{\mu}}_{\alpha'}n_{\nu}\right)\label{der3}
\end{eqnarray}
where $A(\tau)$ and $B(\tau)$ are function of $\tau$ only. For hyperbolic $H^{N}$ spaces, $A=\coth(\tau/a)/a$ and $B=-{\rm csch}(\tau/a)/a$.

Since $\Delta T_{\mu\nu\alpha'\beta'}^{2}(x,x')$ is a symmetric bitensor, one could express it in terms of these basic bitensors. Taking the symmetries of the indices into account, we have
\begin{eqnarray}
\Delta T_{\mu\nu\alpha'\beta'}^{2}(x,x')&=&C_{1}(\tau)n_{\mu}n_{\nu}n_{\alpha'}n_{\beta'}
+C_{2}(\tau)\left(n_{\mu}n_{\nu}g_{\alpha'\beta'}+g_{\mu\nu}n_{\alpha'}n_{\beta'}\right)\nonumber\\ &&\ \ +C_{3}(\tau)\left(n_{\mu}g_{\nu\alpha'}n_{\beta'}+n_{\nu}g_{\mu\alpha'}n_{\beta'}+n_{\mu}g_{\nu\beta'}n_{\alpha'} +n_{\nu}g_{\mu\beta'}n_{\alpha'}\right)\nonumber\\
&&\ \ +C_{4}(\tau)\left(g_{\mu\alpha'}g_{\nu\beta'}+g_{\mu\beta'}g_{\nu\alpha'}\right)+C_{5}(\tau)g_{\mu\nu}g_{\alpha'\beta'}\label{T2bitensor}
\end{eqnarray}
Using the derivatives in Eqs.~(\ref{der1}) to (\ref{der3}), the conservation conditions $\nabla^{\mu}\Delta T_{\mu\nu\alpha'\beta'}^{2}(x,x')=0$ can be expressed as three equations on the coefficients,
\begin{eqnarray}
&&\frac{dC_{1}}{d\tau}+\frac{dC_{2}}{d\tau}-2\frac{dC_{3}}{d\tau}+(N-1)AC_{1}+2BC_{2}-2\left((N-2)A+NB\right)C_{3}=0\nonumber\\
&&\frac{dC_{2}}{d\tau}+\frac{dC_{5}}{d\tau}+(N-1)AC_{2}+2BC_{3}-2(A+B)C_{4}=0\nonumber\\
&&\frac{dC_{3}}{d\tau}-\frac{dC_{4}}{d\tau}+BC_{2}+NAC_{3}-N(A+B)C_{4}=0\label{conservation}
\end{eqnarray}
Moreover, the traceless condition $g^{\mu\nu}\Delta T_{\mu\nu\alpha'\beta'}^{2}(x,x')=0$ can be written as
\begin{eqnarray}
C_{1}+NC_{2}-4C_{3}&=&0\nonumber\\
C_{2}+2C_{4}+NC_{5}&=&0\label{conformal}
\end{eqnarray}

To obtain the five coefficients $C_{1}(\tau)$ to $C_{5}(\tau)$ on $H^{N}$, we just need to evaluate five components of $\Delta T_{\mu\nu\alpha'\beta'}^{2}(x,x')$. Here we shall choose $\Delta T_{\sigma\sigma\sigma'\sigma'}^{2}(x,x')$, $\Delta T_{\sigma\sigma\theta'\theta'}^{2}(x,x')$, $\Delta T_{\sigma\theta\sigma'\theta'}^{2}(x,x')$, $\Delta T_{\theta\theta\phi'_{1}\phi'_{1}}^{2}(x,x')$ and $\Delta T_{\theta\phi_{1}\theta'\phi'_{1}}^{2}(x,x')$. Since $H^{N}$ is homogeneous, the correlator should only depend on the geodesic distance between the two points. Hence, it is possible to simplify the consideration by appropriately choosing $x$ and $x'$. First, we shall set $x$ and $x'$ to have the same angular coordinates $\Omega'\rightarrow\Omega$. After that we shall take the limit $\sigma'\rightarrow 0$. In effect we shall take $x'$ to be at the origin. In addition, in order to keep the consideration simple, we shall only work on the massless minimally coupled case.

As $\Omega'\rightarrow\Omega$, various bitensors can be simplified as follows.
\begin{eqnarray}
\tau(x,x')&=&|\sigma-\sigma'|,\\
n_{\mu}(x,x')&=&\delta_{\mu\sigma},\\
n_{\alpha'}(x,x')&=&-\delta_{\alpha'\sigma'},\\
g_{\sigma\sigma'}(x,x')&=&1,\\
g_{\theta\theta'}(x,x')&=&a^{2}\sinh\left(\frac{\sigma}{a}\right)\sinh\left(\frac{\sigma'}{a}\right),\\
g_{\phi_{1}\phi'_{1}}(x,x')&=&a^{2}\sinh\left(\frac{\sigma}{a}\right)\sinh\left(\frac{\sigma'}{a}\right)\sin^{2}\theta,
\end{eqnarray}
and so on. Note that the non-diagonal elements of $g_{\mu\alpha'}$ vanish in this limit. Using this result the relationship between the various components of $\Delta T_{\mu\nu\alpha'\beta'}^{2}(x,x')$ and the coefficients $C_{i}$ also simplify. Then we have, as $\Omega'\rightarrow\Omega$,
\begin{eqnarray}
\Delta T_{\sigma\sigma\sigma'\sigma'}^{2}(x,x')&=&C_{1}+2C_{2}-4C_{3}+2C_{4}+C_{5},\\
\Delta T_{\sigma\sigma\theta'\theta'}^{2}(x,x')&=&a^{2}\sinh^{2}\left(\frac{\sigma'}{a}\right)(C_{2}+C_{5}),\\
\Delta T_{\sigma\theta\sigma'\theta'}^{2}(x,x')&=&-a^{2}\sinh\left(\frac{\sigma}{a}\right)\sinh\left(\frac{\sigma'}{a}\right)(C_{3}-C_{4}),\\
\Delta T_{\theta\theta\phi'_{1}\phi'_{1}}^{2}(x,x')&=&a^{4}\sinh^{2}\left(\frac{\sigma}{a}\right)\sinh^{2}\left(\frac{\sigma'}{a}\right)\sin^{2}\theta C_{5},\\
\Delta T_{\theta\phi_{1}\theta'\phi'_{1}}^{2}(x,x')&=&a^{4}\sinh^{2}\left(\frac{\sigma}{a}\right)\sinh^{2}\left(\frac{\sigma'}{a}\right)\sin^{2}\theta C_{4},\label{T2comp5}
\end{eqnarray}
The evaluation of the various components of the correlator can be further simplified if we take $\sigma'\rightarrow 0$ as only terms with low values of $l$ and $l'$ will contribute.

To see how the procedure goes, we consider $\Delta T_{\theta\phi_{1}\theta'\phi'_{1}}^{2}(x,x')$ in some details. Using the prescription in Eq.~(\ref{T2finalexp}),
\begin{eqnarray}
&&\Delta T_{\theta\phi_{1}\theta'\phi'_{1}}^{2}(x,x')\nonumber\\
&=&\frac{1}{2}
\int_{0}^{\infty}du u^{\nu}\int_{0}^{\infty}dv v^{\nu}\int_{0}^{\infty}d\kappa\int_{0}^{\infty}d\kappa'\sum_{lml'm'}
e^{-\frac{u}{a^{2}}\left(\kappa_{2}+\frac{(N-1)^{2}}{4}\right)}
e^{-\frac{v}{a^{2}}\left(\kappa_{2}+\frac{(N-1)^{2}}{4}\right)}\nonumber\\
&&\ \ \ \ \ \ \ \ T_{\theta\phi_{1}}[\phi_{\kappa lm}(x),\phi_{\kappa' l'm'}^{*}(x)]T_{\theta'\phi'_{1}}[\phi_{\kappa' l'm'}(x'),\phi_{\kappa lm}^{*}(x')]
\end{eqnarray}
To anticipate that we shall take $\sigma'\rightarrow 0$, we only need to consider the first few values of $l$ and $l'$. For example, when $l=l'=1$, we have
\begin{eqnarray}
T_{\theta\phi_{1}}[\phi_{\kappa 1m}(x),\phi_{\kappa' 1m'}(x)]&=&
-f_{\kappa 1}(\sigma)f_{\kappa'1}(\sigma)\left[\partial_{\theta}Y_{1m'}^{*}(\Omega)\partial_{\phi_{1}}Y_{1m}(\Omega)
+\partial_{\phi_{1}}Y_{1m'}^{*}(\Omega)\partial_{\theta}Y_{1m}(\Omega)\right.\nonumber\\
&&\ \ \ \ \ \ \ \ \ \ \ \ \ \ \ \ \ \ \ \ \ \ \left.-2\xi\nabla_{\theta}\nabla_{\phi_{1}}\left(Y_{1m'}^{*}(\Omega)Y_{1m}(\Omega)\right)\right]
\end{eqnarray}
The summations over $m$ and $m'$ can be carried out using the addition theorem in Eq.~(\ref{addtheorem}). Moreover, in the limit $\Omega'\rightarrow\Omega$, we have
\begin{eqnarray}
\left.\partial_{\theta}\partial_{\theta'}(\Omega\cdot\Omega')\right|_{\Omega'\rightarrow\Omega}&=&1\\
\left.\partial_{\phi_{1}}\partial_{\phi'_{1}}(\Omega\cdot\Omega')\right|_{\Omega'\rightarrow\Omega}&=&\sin^{2}\theta\\
\left.\partial_{\phi_{1}}\partial_{\phi'_{1}}(\Omega\cdot\Omega')^{2}\right|_{\Omega'\rightarrow\Omega}&=&2\sin^{2}\theta\\
\left.\partial_{\theta}\partial_{\phi_{1}}\partial_{\phi'_{1}}(\Omega\cdot\Omega')^{2}\right|_{\Omega'\rightarrow\Omega}&=&2\sin\theta\cos\theta\\
\left.\partial_{\theta}\partial_{\theta'}\partial_{\phi_{1}}\partial_{\phi'_{1}}(\Omega\cdot\Omega')^{2}\right|_{\Omega'\rightarrow\Omega}&=&2
\end{eqnarray}
and the summations can be simplified to
\begin{eqnarray}
&&\left.\sum_{mm'}T_{\theta\phi_{1}}[\phi_{\kappa 1m}(x),\phi_{\kappa' 1m'}^{*}(x)]T_{\theta'\phi'_{1}}[\phi_{\kappa' 1m'}(x'),\phi_{\kappa 1m}^{*}(x')]\right|_{\Omega'\rightarrow\Omega}\nonumber\\
&=&\frac{2\Gamma^{2}\!\left(\frac{N}{2}+1\right)}{\pi^{N}}\sin^{2}\theta \left[f_{\kappa 1}(\sigma)f_{\kappa'1}(\sigma)f_{\kappa 1}(\sigma')f_{\kappa'1}(\sigma')\right]
\end{eqnarray}
Furthermore, as $\sigma'\rightarrow 0$,
\begin{eqnarray}
&&\left.\sum_{mm'}T_{\theta\phi_{1}}[\phi_{\kappa 1m}(x),\phi_{\kappa' 1m'}^{*}(x)]T_{\theta'\phi'_{1}}[\phi_{\kappa' 1m'}(x'),\phi_{\kappa 1m}^{*}(x')]\right|_{\Omega'\rightarrow\Omega,\sigma'\rightarrow 0}\nonumber\\
&=&\frac{2}{(2\pi)^{N}}\sinh^{2}\left(\frac{\sigma'}{a}\right)\sin^{2}\theta\left[c_{1}(\kappa)f_{\kappa 1}(\sigma)c_{1}(\kappa')f_{\kappa'1}(\sigma)\right]
\end{eqnarray}
where we see that $\sinh^{2}(\sigma'/a)$ is the leading behavior as $\sigma'\rightarrow 0$. This behavior conforms with that of $\Delta T_{\theta\phi_{1}\theta'\phi'_{1}}^{2}(x,x')$ in Eq.~(\ref{T2comp5}) and that is the only term with this leading behavior.
Comparing with Eq.(\ref{T2comp5}), we can extract the coefficient $C_{4}$.
\begin{eqnarray}
C_{4}(\sigma)=\frac{1}{(2\pi)^{N}a^{4}\left(\sinh\frac{\sigma}{a}\right)^{2}}\left(I_{1}^{(0)}\right)^{2}
\end{eqnarray}
where we have defined the integral,
\begin{eqnarray}
I_{l}^{(i)}=\int_{0}^{\infty}du u^{\nu}\int_{0}^{\infty}d\kappa\kappa^{2i} e^{-\frac{u}{a^{2}}\left(\kappa^{2}+\frac{(N-1)^2}{4}\right)}c_{l}(\kappa)f_{\kappa l}(\sigma)\label{Iintegral}
\end{eqnarray}

After some lengthy calculations similar to that above, we have
\begin{eqnarray}
C_{1}&=&-\frac{1}{(2\pi)^{N}a^{4}}\left[\frac{(N-1)^{2}a^{2}}{4}\left(\partial_{\sigma}I_{0}^{(0)}\right)^{2}
+a^{2}\left(\partial_{\sigma}I_{0}^{(0)}\right)\left(\partial_{\sigma}I_{0}^{(1)}\right)\right]\nonumber\\
&&+\frac{1}{(2\pi)^{N}a^{4}}\left[2a^{2}\left(\partial_{\sigma}I_{1}^{(0)}\right)^{2}
+I_{1}^{(0)}I_{1}^{(1)}-\left(\frac{2a}{\sinh\frac{\sigma}{a}}
\partial_{\sigma}-\frac{(N-1)^{2}}{4}-\frac{2}{\left(\sinh\frac{\sigma}{a}\right)^{2}}\right)
\left(I_{1}^{(0)}\right)^{2}\right]\nonumber\\ \\
C_{2}&=&\frac{1}{(2\pi)^{N}a^{4}}\left[\frac{(N-1)^{2}a^{2}}{4}\left(\partial_{\sigma}I_{0}^{(0)}\right)^{2}
+a^{2}\left(\partial_{\sigma}I_{0}^{(0)}\right)\left(\partial_{\sigma}I_{0}^{(1)}\right)\right]\nonumber\\
&&\ \ \ +\frac{1}{(2\pi)^{N}a^{4}}\left[\frac{1}{\left(\sinh\frac{\sigma}{a}\right)^{2}}\left(I_{1}^{(0)}\right)^{2}
-a^{2}\left(\partial_{\sigma}I_{1}^{(0)}\right)^{2}\right]\label{Coeff1}\\
C_{3}&=&-\frac{1}{2(2\pi)^{N}a^{4}\left(\sinh\frac{\sigma}{a}\right)^{2}}
\left[a\left(\sinh\frac{\sigma}{a}\right)\partial_{\sigma}-2\right]
\left(I_{1}^{(0)}\right)^{2}\\
C_{4}&=&\frac{1}{(2\pi)^{N}a^{4}\left(\sinh\frac{\sigma}{a}\right)^{2}}
\left(I_{1}^{(0)}\right)^{2}\\
C_{5}&=&\frac{1}{2(2\pi)^{N}a^{4}}\Bigg[-a^{2}\left(\partial_{\sigma}I_{0}^{(0)}\right)
\left(\partial_{\sigma}I_{0}^{(2)}\right)-\frac{a^{2}(N-1)^{2}}{4}
\left(\partial_{\sigma}I_{0}^{(0)}\right)^{2}\nonumber\\
&&\ \ \ \ \ \ \ \ \ \ \ \ \ \ \ \ \ \ \ \ +\left(I_{0}^{(1)}\right)^{2}+\frac{(N-1)^{2}}{2}
\left(I_{0}^{(0)}I_{0}^{(1)}\right)+\frac{(N-1)^{4}}{16}
\left(I_{0}^{(0)}\right)^{2}\Bigg]\nonumber\\
&&+\frac{1}{2(2\pi)^{N}a^{4}}\left[a^{2}
\left(\partial_{\sigma}I_{1}^{(0)}\right)^{2}
-I_{1}^{(0)}I_{1}^{(1)}+\left(
-\frac{(N-1)^{2}}{4}+\frac{N-5}{\left(\sinh\frac{\sigma}{a}\right)^{2}}\right)
\left(I_{1}^{(0)}\right)^{2}\right]\nonumber\\ \label{Coeff5}
\end{eqnarray}

Although we have succeeded in expressing the coefficients in terms of products of integrals involving the associated Legendre functions, the integrals cannot be simplified to known functions. In this form it is hard to check the corresponding conservation conditions. Therefore, we shall consider in the next section the asymptotic behaviors of the above coefficients in the small and large geodesic distance limits. In so doing we could have a better understanding of the correlators and we could also check the conservation and the traceless conditions explicitly for various dimensions.

\section{Small and large geodesic distance limits in AdS for different dimensions}

In this section we explore the small and the large geodesic distance limits of the coefficients $C_{1}$ to $C_{5}$ in Euclidean $AdS^N$ for arbitrary $N$.  As we have mentioned above, it is possible to check the conservation and the traceless conditions explicitly for various dimensions in the small distance limit. In addition, the small distance limit also indicates the divergent behavior of the correlators in the coincident limit as $\sigma\rightarrow 0$. This type of expansion has been explored in \cite{OsbSho00} in relation to the generalization of the $c$-theorem \cite{Zam86} from two to higher dimensions. A general discussion has also been given in \cite{OsbSho00} on the stress-energy correlators of conformally coupled scalar and fermion fields in constant curvature spaces.

For the small distance limit, we need to consider the integral $I_{l}^{(i)}(N)$ in Eq.~(\ref{Iintegral}) in more detail. First we shall use the integral representation of the Legendre function,
\begin{equation}
P_{-\frac{1}{2}+i\kappa}^{1-l-\frac{N}{2}}\left(\cosh\frac{\sigma}{a}\right)=
\frac{\sqrt{2}\left(\sinh\frac{\sigma}{a}\right)^{1-l-\frac{N}{2}}}{\sqrt{\pi}\Gamma\left(-\frac{1}{2}+l+\frac{N}{2}\right)}
\left(\frac{\sigma}{a}\right)\int_{0}^{1}dt\ \!\cos\left(\frac{\kappa\sigma t}{a}\right)\left(\cosh\frac{\sigma}{a}-\cosh\frac{\sigma t}{a}\right)^{l+\frac{N}{2}-\frac{3}{2}}
\end{equation}
Since the normalization constant is given by
\begin{equation}
|c_{l}(\kappa)|^{2}=\left[\kappa^{2}+\left(\rho_{N}+l-1\right)^{2}\right]\left[\kappa^{2}+\left(\rho_{N}+l-2\right)^{2}\right]
\cdots\left[\kappa^{2}+\left(\rho_{N}\right)^{2}\right]|c_{0}(\kappa)|^{2}
\end{equation}
and for even dimensions, $|c_{0}(\kappa)|^{2}$ is given by Eq.~(\ref{c0even}),
we see that if we want to extract the leading contribution of the integrals as $\sigma\rightarrow 0$, we need to deal with integrals of the general form
\begin{eqnarray}
&&\int_{0}^{1}dt\ \!\left(\cosh\frac{\sigma}{a}-\cosh\frac{\sigma t}{a}\right)^{l+\frac{N}{2}-\frac{3}{2}}\int_{0}^{\infty}du\ \! u^{\nu}\int_{0}^{\infty}d\kappa\ \! \kappa^{2n+2i+1}\left(\cos\frac{\kappa\sigma t}{a}\right)\ \! e^{-\frac{u}{a^{2}}\left(\kappa^{2}+\frac{(N-1)^2}{4}\right)}\nonumber\\
&=&\int_{0}^{1}dt\ \!\left(\cosh\frac{\sigma}{a}-\cosh\frac{\sigma t}{a}\right)^{l+\frac{N}{2}-\frac{3}{2}}
\int_{0}^{\infty}d\kappa\frac{a^{2(\nu+1)}\Gamma(\nu+1)\kappa^{2n+2i+1}\left(\cos\frac{\kappa\sigma t}{a}\right)}{\left(\kappa^{2}+\frac{(N-1)^2}{4}\right)^{\nu+1}}
\end{eqnarray}
where $n$ is an integer. This is finite for sufficiently large values of $\nu$. In this analytic continuation procedure, we can represent $\kappa^{2n+2i}$ by derivatives on the cosine function. Then the above integral becomes
\begin{equation}
\int_{0}^{1}dt\ \!\left(\cosh\frac{\sigma}{a}-\cosh\frac{\sigma t}{a}\right)^{l+\frac{N}{2}-\frac{3}{2}}
(-1)^{n+i}\left(\frac{\partial}{\partial\eta}\right)^{2n+2i}\int_{0}^{\infty}d\kappa\left[\frac{a^{2}\kappa\ \!\cos\eta\kappa}{\kappa^{2}+\frac{(N-1)^2}{4}}\right]
\end{equation}
where $\eta=\sigma t/a$. Since the integration over $\kappa$ is now finite even when $\nu\rightarrow 0$, we have taken that limit.

To see how it goes explicitly, we consider the case with $l=1$, $N=4$ and $n=i=1$. Then, expanding in powers of $\sigma/a$,
\begin{eqnarray}
&&\left(\cosh\frac{\sigma}{a}-\cosh\frac{\sigma t}{a}\right)^{3/2}\left(\frac{\partial}{\partial\eta}\right)^{4}\int_{0}^{\infty}d\kappa\left(\frac{a^{2}\kappa\ \!\cos\eta\kappa}{\kappa^{2}+\frac{9}{4}}\right)\nonumber\\
&=&a^{2}\left(\frac{a}{\sigma}\right)\left(\frac{3}{\sqrt{2}}\right)(1-t^2)^{3/2}t^{-4}+
a^{2}\left(\frac{\sigma}{a}\right)\left(\frac{3}{8\sqrt{2}}\right)(1-t^{2})^{3/2}\left[t^{-4}+4t^{-2}\right]\nonumber\\
&&\ \ \ +a^{2}\left(\frac{\sigma}{a}\right)^{3}\left(\frac{1}{640\sqrt{2}}\right)(1-t^{2})^{3/2}\Bigg\{13t^{-4}+108t^{-2}
+103
\left.-810\left[2\gamma+\ \!{\rm ln}\left[\frac{9t}{4}\left(\frac{\sigma}{a}\right)^{2}\right]\right]\right\}\nonumber\\
\end{eqnarray}
where we have put $a^{2}b=9/4$ for $N=4$. Using the formula,
\begin{equation}
\int_{0}^{1}dt\ \!t^{\alpha}(1-t^2)^{\beta}=\frac{\Gamma\left(\frac{1+\alpha}{2}\right)
\Gamma\left(1+\beta\right)}{2\Gamma\left(\frac{3}{2}+\frac{\alpha}{2}+\beta\right)}
\end{equation}
the integration over $t$ can be performed giving
\begin{eqnarray}
&&\int_{0}^{1}dt\ \!\left(\cosh\frac{\sigma}{a}-\cosh\frac{\sigma t}{a}\right)^{3/2}\left(\frac{\partial}{\partial\eta}\right)^{4}\int_{0}^{\infty}d\kappa\left(\frac{\kappa\ \!\cos\eta\kappa}{\kappa^{2}+\frac{9}{4}}\right)\nonumber\\
&=&a^{2}\left(\frac{a}{\sigma}\right)\left(\frac{3\pi}{2\sqrt{2}}\right)
-a^{2}\left(\frac{\sigma}{a}\right)\left(\frac{30\pi}{32\sqrt{2}}\right)
-a^{2}\left(\frac{\sigma}{a}\right)^{3}\left(\frac{\pi}{10240\sqrt{2}}\right)\times\nonumber\\
&&\ \ \left\{883-1215
\left[3-4\gamma-4\ \!{\rm ln}\left(\frac{3}{4}\right)-4\ \!{\rm ln}\left(\frac{\sigma}{a}\right)\right]\right\}
\end{eqnarray}

Since the coefficients $C_{1}$ to $C_{5}$ are expressed in terms of the integrals $I_{l}^{(i)}(N)$, the expansion above can be used to obtain the corresponding small geodesic distance limits of the coefficients.
For $N=4$,
\begin{eqnarray}
C_{1}(4)&=&\frac{8}{\pi^{4}\sigma^{8}}\left(1-\frac{7\sigma^{2}}{12a^{2}}+\cdots\right)\nonumber\\
C_{2}(4)&=&-\frac{2}{\pi^{4}\sigma^{8}}\left(1+\frac{19\sigma^{2}}{6a^{2}}+\cdots\right)\nonumber\\
C_{3}(4)&=&\frac{1}{\pi^{4}\sigma^{8}}\left(1-\frac{23\sigma^{2}}{24a^{2}}+\cdots\right)\nonumber\\
C_{4}(4)&=&\frac{1}{4\pi^{4}\sigma^{8}}\left(1-\frac{4\sigma^{2}}{3a^{2}}+\cdots\right)\nonumber\\
C_{5}(4)&=&\frac{1}{\pi^{4}\sigma^{8}}\left(1-\frac{5\sigma^{2}}{24a^{2}}+\cdots\right)\nonumber\\ \label{N4coeff}
\end{eqnarray}
One can then check that the conservation conditions in Eq.~(\ref{conservation}) are satisfied to the order indicated.

With the expressions in the small distance limit of the coefficients $C$'s, it is possible to make some comparison with our results to that in \cite{PRVdS}. For simplicity, we just consider the leading behavior of these coefficients. In \cite{PRVdS}, these coefficients, named $P$, $Q$, $R$, $S$, and $T$ there, are expressed in terms of the Wightman function (in de Sitter spacetime),
\begin{eqnarray}
G(\sigma)=c_{m,N}F\left(h_{+},h_{-};\frac{N}{2};\frac{1+\cos\sigma/a}{2}\right)
\end{eqnarray}
where $F(\alpha,\beta;\gamma;z)$ is the hypergeometric function and
\begin{eqnarray}
h_{\pm}=\frac{1}{2}\left[(N-1)\pm\sqrt{(N-1)^{2}-4m^{2}a^{2}}\right]\ \ ;\ \ c_{m,N}=\frac{\Gamma(h_{+})\Gamma(h_{-})}{(4\pi a)^{N/2}\Gamma(N/2)}
\end{eqnarray}
In the small geodesic distance limit, $\sigma\rightarrow 0$,
\begin{eqnarray}
G(\sigma)=\frac{\Gamma\left(\frac{N}{2}-1\right)}{(4\pi)^{N/2}}\left(\frac{2}{\sigma}\right)^{N-2}+\cdots.
\end{eqnarray}
The leading behavior of the coefficients are then
\begin{eqnarray}
C_{1}(N)&=&\frac{N^{2}\Gamma^{2}(N/2)}{2\pi^{N}\sigma^{2N}}+\cdots\nonumber\\
C_{2}(N)&=&-\frac{N(N-2)\Gamma^{2}(N/2)}{4\pi^{N}\sigma^{2N}}+\cdots\nonumber\\
C_{3}(N)&=&\frac{N\Gamma^{2}(N/2)}{4\pi^{N}\sigma^{2N}}+\cdots\nonumber\\
C_{4}(N)&=&\frac{\Gamma^{2}(N/2)}{4\pi^{N}\sigma^{2N}}+\cdots\nonumber\\
C_{5}(N)&=&\frac{(N^{2}-N-4)\Gamma^{2}(N/2)}{8\pi^{N}\sigma^{2N}}+\cdots\label{PRVdScoeff}
\end{eqnarray}
These agree with our results in Eq.~(\ref{N4coeff}) when one puts $N=4$. It is also easy to show that the coefficients in Eqs.~(\ref{Coeff1}) to (\ref{Coeff5}) indeed have the same leading behavior as above.

To consider the large geodesic distance limit for the various coefficients, we go back to the integral $I_{l}^{(i)}$ in Eq.~(\ref{Iintegral}). First,
\begin{equation}
f_{\kappa l}(\sigma)=c_{l}(\kappa)\left(\sinh\frac{\sigma}{a}\right)^{1-\frac{N}{2}}P_{-\frac{1}{2}+i\kappa}^{1-l-\frac{N}{2}}\left(\cosh\frac{\sigma}{a}\right)
\end{equation}
and from the asymptotic behavior as $\sigma\rightarrow\infty$ of the associated Legendre function, we have
\begin{equation}
P_{-\frac{1}{2}+i\kappa}^{1-l-\frac{N}{2}}\left(\cosh\frac{\sigma}{a}\right)\sim e^{-\frac{\sigma}{2a}}e^{i\frac{\kappa\sigma}{a}}
\end{equation}
The integrations over $u$ and $\kappa$ will only give powers of $\sigma$. Hence, the leading asymptotic behavior of $I_{l}^{(i)}$, independently of the parameters $l$ and $i$, is
\begin{equation}
I_{l}^{(i)}\sim e^{-(N-1)\sigma/2a}
\end{equation}
Using this result we can estimate the asymptotic behavior of the various coefficients in Eqs.~(\ref{Coeff1}) to (\ref{Coeff5}),
\begin{eqnarray}
C_{1},C_{2},C_{5}&\sim& e^{-(N-1)\sigma/a}\nonumber\\
C_{3}&\sim&e^{-N\sigma/a}\nonumber\\
C_{4}&\sim&e^{-(N+1)\sigma/a}\label{large}
\end{eqnarray}
This behavior is in fact independent of the mass of the field. We can see that an intrinsic infrared cutoff scale is provided by the radius $a$ of the background spacetime.

This large geodesic distance limit can also be compared with the result in \cite{PRVdS}. Since there the authors considered the de Sitter spacetime, one could analytic continue their result by taking
\begin{equation}
Z=\cos\left(\frac{\sigma}{a}\right)\rightarrow\cosh\left(\frac{\sigma}{a}\right)\sim e^{\sigma/a}
\end{equation}
Then our result in Eq.~(\ref{large}) indeed agrees with that in \cite{PRVdS}.

\section{Conclusions and Discussions}

In this paper we have calculated the correlators of the stress-energy tensor, or the noise kernel in the theory of stochastic gravity in $AdS^{N}$ spacetimes. The method we have used is the generalized zeta-function regularization procedure devised in \cite{PH97}. The zeta-function regularization method was originally used to deal mainly with one loop effective action and the corresponding operators associated with it, notably the expectation value of the stress-energy tensor. The method introduced in \cite{PH97} enables one to deal also with operator correlation functions. To understand the correlations more closely and also to facilitate further applications, we have developed their small and long geodesic distance limits. Particularly in the small geodesic distance limit, we could verify explicitly that the correlators satisfy the conservation equation.

We have compared our results with that in \cite{PRVdS}. The agreement shows that the zeta-function method used here is an useful alternative to evaluate correlators of operators. We are confident that this method can be used in other applications which are also of significant interest to us. As discussed in the Introduction, we would like ultimately to explore the physics in the AdS-black hole spacetimes. In order to do that we need to extend our consideration to first the finite temperature case, and then to the black hole case. These cases will be considered in our subsequent works.

\ack

HTC is supported in part by the National Science Council of the Republic of China under the Grant NSC 99-2112-M-032-003-MY3, and by the National Center for Theoretical Sciences. BLH is supported in part by the National Science Foundation under grants PHY-0601550 and PHY-0801368 to the University of Maryland.

\end{document}